# Astronomy of Cholanaikkan tribe of Kerala

M N Vahia[1], V S Ramachandran[2], Jayant Gangopadhyay[2], Justin Joseph[2]
1 Tata Institute of Fundamental Research, Mumbai
2 Regional Science Centre & Planetarium, Kozhikode

**Abstract**

Cholanaikkans are a diminishing tribe of India. With a population of less than 200 members, this tribe living in the reserved forests about 80 km from Kozhikode, it is one of the most isolated tribes. A programme of the Government of Kerala brings some of them to Kozhikode once a year. We studied various aspects of the tribe during such a visit in 2016. We report their science and technology.

1. **Background:**

*Cholanaikkans* are a small tribe with a total population of less than 200 persons, a fraction of whom live in caves, the rest in temporary self-built structures (see e.g. en.wikipedia.org/wiki/Cholanaikkan). Their region of habitation is about 90 km from the city of Kozhikode (Calicut) in a reserved forest[*]. *A detailed description of their life is available at (2, 6)*. As per the government arrangements, no one is allowed to go to within 20 km of their habitation. Their forest area is completely protected and outsiders are not even allowed to scavenge for forest produce or contact the tribal communities. The only contact with the people is done through a weekly (every Wednesday) meeting on the banks of a river between the last checkpoint and their residence, about 10 km from the last permitted check point. This spot is at the foot of their mountain at a distance of about 8 km. The meeting occurs around noon and last for about an hour and a half. There they are fed and given ration of rice etc. that would last them for a week.

The *Cholanaikkans* (*coolanaaykkan*) are called the Cavemen of Kerala (5). This primitive hill tribe inhabits the forests in the Nilambur Valley of Malappuram district of Kerala. The *Cholanaikkan* habitations are on the banks of the rivers in this valley. They still live in the rock shelters called *aale*. The people call themselves as *Colekkaru* (*coolekkaarŭ*) while others refer to them as *coolanaykkar*. The caves or rock shelters of these people have names. They distinguish between two persons having the same name, by adding their caves name before their proper name. They affix *–nu* to their names. The existence of this tribe was unknown until the 1971 census. It is only after 1977 that these people began to receive considerable attention at the national level. The total population of this community is 176. In some places they intermingle with *Kattunaickan* and *Pathinayakkan* tribes. A sample of members of the tribe are given in Figure 1.

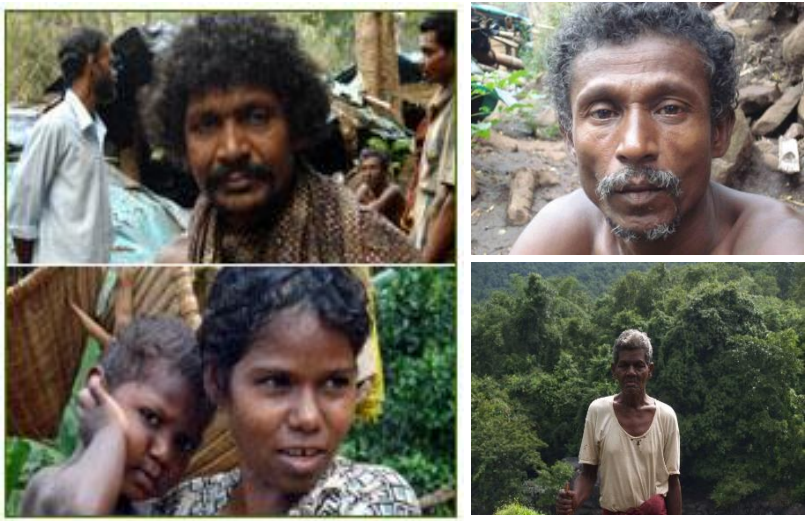

*Figure 1 Faces of Cholannaikans. Some images are taken from (1).*

The present report is based on an annual programme arranged by the Regional Science Centre & Planetarium, Kozhikode in association with Kerala Forest Department; to bring some of them for a city tour to Kozhikode and to the local science centre once a year. The present encounter occurred on Mach 3, 2016 and included about 50 Cholanaikkars, many venturing out of their house for the first time in their lives(figure 2). The most outstanding features of the community are listed below.

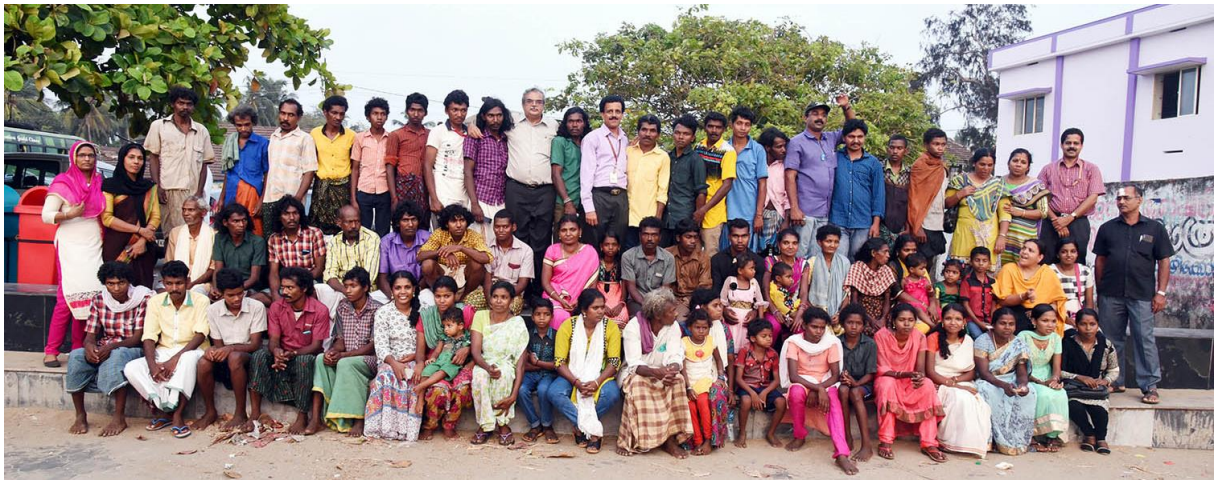

*Figure 2: Photograph of Cholanaikkans with Planetarium staff and volunteers at beach in Kozhikode on 3 March 2016*

## 2. About the tribe:

The Cholanaikkan tribe is classified as Primitive Tribal Group by Government of India (6). They call themselves as *Malanaikan* or *Sholanaikan*. *Shola* or *chola* means deep thicket in the forest and *naikan* means king (7). They are called *Cholanaikkan* because they inhabit the interior forests. They are said to have migrated from Mysore forests. The reason for their settling in the Nilambur forests has two versions. One is that they failed in a battle and had to hide deep in the forest. The second version is that they were displaced by flood and they sought shelter in the forests. However the only credential to substantiate the former version is an antique sword which the community still holds on. The total population of the community is 56 families comprising of a *total population* of 176 persons. This is down from about 360 people in 1991 census. They are divided into smaller groups called *Jenmam.* They have no fixed dwellings but prefer to live close to water sources. They live in rock shelters called *Kallulai (Kallu* means rock, *aalai* means 'cave') or in open campsites made of leaves. They are found in groups consisting of 2 to 7 primary families. Each group is called a *Chemmam*. They are the only cave dwelling community in India. Each cave of the community has a name. And people with same names are distinguished by suffixing their name with that of their caves.The dwellers of rock shelters are called *Mannalar*. *The group must have been much larger in the past as they remember 4 different identities of Arnadan (largest with 148 people), Kattunayikkar, Cholanayikkar and Paniyar.* The community lives essentially by scavenging the forest.They do not cultivate due to the problem of elephants trampling over their produce. Their day starts with sunrise and ends with sunset – they do not use fire to extend the day. Theylive in a protected forest with elephants and other wild animals including occasional sighting of tigers. Antelopes are very common. Their typical house is shown in figure 3.

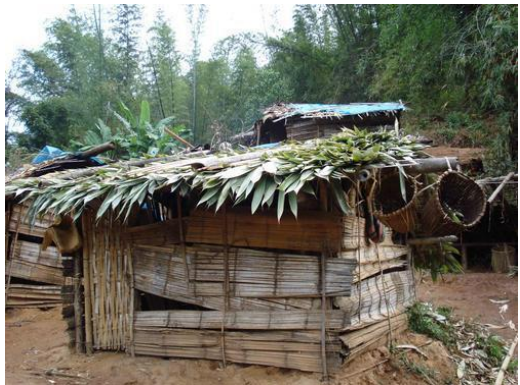

*Figure 1: House of Cholanaikkans, from (11)*

*Until about 30 years ago (till mid to late 1980's) they wore no clothes and their bodies were not covered in any way. They now have clothing provided by government and other charities.* They have meagre wealth and no weapons. It seems that they use no weapons – *they now use iron tools provided by the government but it is not clear if they used metal tools (other than wood) in the past*. They occasionally come with surplus bamboo baskets and other forest produce which they sell to the government agency. The community has no use of this but the youngsters occasionally going to school etc. use this money. They speak a language which is weakly called Dravidian but it is not directly related to any of the modern Dravidian languages – the sound of their talking gives a whiff of similarity to Malyalam, Tamil and Telugu but direct association of words is not seen. The standard claim is that the language is a mix of Malyalam and Kanada. A detailed study of the language (5 see section 10) suggests that the language is Dravidian. However it does seem to have a lot root words of many of the modern words in the Dravidian languages. Their names also suggest local words and do not carry any names from Hindu mythologies or other more prevalent names suggesting long period of isolation.

3. **About the people:**

They are generally of short stature about 1.6 meters (5 feet 5 inches) with well-built sturdy bodies. The complexion varies from dark to light brown. The faces are round or oval with depressed nasal root, their bridge being medium and the profile straight, lips are thin to the medium, hair tends to be curly. They tend to have some features which are more African like than that of the common population. No genetic profile is available. It is likely that they are Austro Asian in origin. The group that we met was a good mix of the young and the old but their age is impossible to determine. The absence of infants in the group must almost certainly be data bias as the people with infants may not want to travel. *Cholanaikkans* are animists and worship 'ancestral spirits'. They don't have any idols or images but go behind a big tree and whisper something. Scavenging and fishing are done by both the genders but men play the primary role with women assisting more in gathering and transportation. Household chores such as storing food grains, cleaning, drying, cooking, sharing and entertaining guests are done by the women. In the absence of a woman a male member may do the cooking and serving (3).They primarily eat plants, leaves, fruits, mushrooms, seeds and tubers but can also include fish, birds, rabbits, pigs, wild buffalos, monkeys and turtles. However, they do not use milk at all. (3). In general however, they tend to be vegetarian. Now they augment it with the ration that the Government gives them. The way they entrap monkey is quite peculiar. It is a collective venture. They chase the monkey and track it towards a tree, which stands separated in the forest. Once they make sure that there is no way out for the monkey to get away they would cut down the tree to catch the animal. The *Cholanaikkans* tame animals and birds only as a hobby. Though they tame cows and birds like parrot, hen and hornbills they never use them for milk, egg or meat. Thus, domestication and using them not a part of their culture. *Cholanaikkans* abide by a set of rules framed by their ancestors. The Chief of the community is called the *Mooppan*. He has the authority to resolve disputes among their members. The authority and power of the *Mooppan* is transferred to the next generation symbolically by conferring his inherited Staff of Power to whomever he wishes to transfer. And invariably it would be to his son if he has one. The Staff of Power is called the *Chemmathadi*. They are governed by a council of elders (called *Jenmekkaram*) which controls the activities of the group.

4. **Their worship**

For worship, they take an undressed stone with a curved top (shaped like the head of a snake), keep it under a large tree and worship it. But have no great expectations from their god. The tree is called the '*Dheivamaram*' which means tree of god. They do have a concept of ancestor worship – but do have more broad based spirit worship. They worship divine power and spirits. They worship god of the jungle (*MalalliDiavam*), metal image of a tiger (*Uliuruvu*), and images of Ox (*Kalaiuruvu*) and snake god. (3).They refer to Sun as *Nyaram* or *Dinga* and Moon as *Thinkam*.They regard Sun with reverence and bury their dead with the head to the East. They celebrate their own forest festivals when they percussion musical instruments of their own creation. One of their traditional feasts which can be compared with the harvest festival of farming communities is *'Dheivaoottu'*(meaning - feeding the deity). This festival is during the honey harvesting season invariably in April. People offer a part of the honey to their god.They do not have rituals for birth. However, when a child is born in a *Cholanaikkan* family they lay the child on the ground in east-west direction (the head towards the east). Just as the 'east' marks the beginning of the day, it represents the dawn of the individual's life.

5. **Marriage customs**

They only marry within the community and avoid all contact with the outside world. They prefer to have a bride from another subgroup of their own group and prefer to marry mother's brother's daughter/son or father's sister's daughter/son as a companion (3). It is customary for a young man to find his mate. If they wish to continue their relations, the husband (*gunda*) and the wife (*ennu*) they allow themselves to be seen by the girl's parents or the local chieftain. Thus the relationship is recognised as a marriage companion (3). In the event of too many men wanting the same bride, the community chose the groom through certain competitions. It could be like climbing up a tree with many honeycombs and bring down maximum honey. And in the event of tie the competitors had to prove their mettle venturing into wrestling! The winner marries marry the girl. The dowry system is prevailing in the community. But their system of dowry called *Mothalana*. Itis rather different from conventional dowry system. There are two types of *Mothalanas*. One is given at the time of marriage and the other at the time of the death of husband. At the time of marriage *Mothalana* has to be given to the bride by the groom's family. At the time of the death of a husband she is entitled for *Mothalana* from the siblings of her husband. Their traditions and customs have paramount importance to for safety and welfare of women in the community. A girl may also be abducted by the boy and such a capture is called *'edippiyodu'* resulting in a sexual union (*oppamaladu)*, and if they allow themselves to be seen later to be living together, the marriage is considered complete companion (3).They do not have any ceremony in connection with the wedding companion (3).In the present day, marriages are arranged by elders. The marriages are patriarchal – the bride goes to the groom's family but each couple has a separate house companion (3). Polygamy is generally discouraged, but a man in the community may have a maximum of 3 wives provided he can demonstrate a capability to sustain them. On passing away of a spouse, the partner (male or female) is allowed to remarry. A girl may be married at an age as young as 12 to 13 years as rejecting a proposal may result in not being proposed to at all (2).

**6. Efforts to integrate them.**

In order to assist them, the government of Kerala has a weekly interaction programme every Wednesday. During these meetings, they are given weekly ration of rice, wheat, oil, potatoes, spices and onions. During such interactions the Cholanaikkans may also sell their items such as honey, baskets etc.(7).The state government has built about thirty houses for them by the banks of river Karimpuzha. Further, government also has provided a one teacher residential school (called the Indira Gandhi Modern Residential school at Karulai) for them. However, all but two houses have been abandoned. No takers for these permanent dwellings (except two houses) since they prefer to stay in their original habitations. About 20 children have passed through this school at some stage. Children coming to the residential school (at Karulai) typically last one semester only and the dropout at the end of one semester is very significant(2, 8). So while some youngsters can speak broken Malayalam, the literacy rate is very poor and well below the official claim of 16%. However, there is a general feeling of the need to be educated. Only one youngster (Vinod) from the family has left the village to pursue studies – his initial education was in Ernaculam and is studying for B Com in Pathanamthitta. He was not amongst the visitors. His ambition in life is to become and IAS officer. Three of the 50 families claim to have someone with basic literacy amongst them.*However, the mobile phone penetration is 98%.* From this it may be inferred that they have abstract concept of numbers as a tool for counting, or they may understand it as a string of symbols with or without order. While men and women of the group tend to move and sit separately, there is, in general no gender separation nor gender suppression. The women are quite happy to take lead in activities. During this visit, many in the community saw the sea for the first time. All of them refused to go in, excluding about 10 youngsters who went in to wet their feet. However, egged on by the female volunteers, the women were quite happy to play common beach games while the men continued to cluster and watch.

## 7. Recent history and some observations

Recently they lost their Chief (Moopan) who passed away. The people were afraid of him since he knew black magic and most people avoided him. Now the Chief is a man named Konkan. He has a sister named Vellan. She is married to Panapooyam, whose main profession is selling 'pantham' (*fire torches*). They have five children who are all being educated. A person called Balan who has studied till class 12 works as a forest guard and provides a link to the community. His stated his age as 34 years. During the meeting with the ration distributing people, the community shows no really interest in communicating or mixing with outsiders, preferring to keep the contact to a minimum. Not being involved with others all their lives, *their cultural expressions were also different* – none for example clapped at magic show of the 3D show even though others in the audience did. They were generally unimpressed (or non-comprehending) of what they saw – it just amused them and occasionally they smiled to themselves but there were no animated discussions or excited gesturing about what they had experienced. However, when asked to talk amongst themselves in their own language in front of a video camera, they were quite excited and merrily talked, but by and large men and women had separate discussion groups and there was little cross talk between males and females. Members often remained separate by gender to the extent that it was impossible to even guess who was married to whom – neither the spouses nor the children of a family form separate subgroups. The segregation was by age and gender. There were very few exceptions. One young child, confused and scared searched for security with an elder, presumably her father who provided protection without any serious communication between them. She just came and sat on the father's lap and then they moved together. Hardly any word was exchanged between them. In recent times, due to decline in female population they have married women from another equally unknown but relatively larger tribe called '*Kattunaikkans*'. All through the trip, the group remained unconcerned about the outsiders – showing neither anxiety nor interest and only occasionally looked around the galleries at the science centre. They exhibited no interest in communicating with the hosts. Even when felicitated, they remained indifferent. Some youngsters who could speak Malayalam and had probably seen more of the world were an exception. The impression of the group is that they are largely pacifistic and not given to violence. The children seem remarkably unconstrained and seem to be growing up in a relatively unstressed environment. This is confirmed by a more detailed description of their life *in situ* (2). However, no one ill-behaved, neither displayed curiosity about exhibits to the level of being destructive. At the same time, they were willing to experiment with some of the strange demonstrative artefacts of the Science Centre. While eating, appear to satisfy hunger rather than relishing the taste – they often ate each item in the food – rice, dal and chicken separately. However, there was little wastage of food and they left their eating place clean unlike the discarded material left behind by the volunteers who were with them.

## 8. Their science:

During their visit to the Science Centre we had an opportunity to study their science as well as study their response to various displays of scientific demonstrations at the museum. In figure 4 we have given a photograph of a grandfather – granddaughter peeking through a multiple reflection display at the museum. In general, their response to such displays was of mild amusement and they seemed to have little desire to acquire any of the gadgets for themselves. In figure 5 we have reproduced from [11] an example of their use of tools to divert water though use of bamboo shoots.

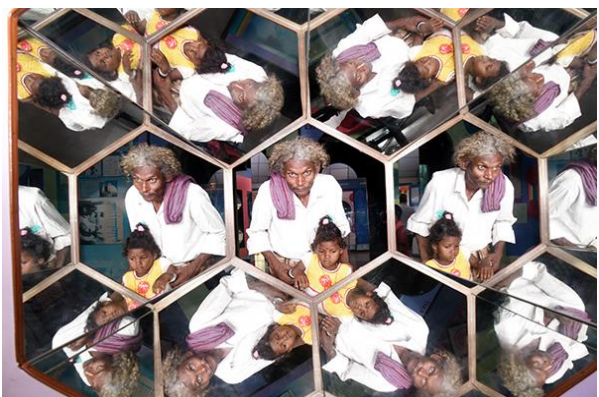

*Figure 2: The elder of the tribe looking into a kaleidoscope at the Science Centre.*

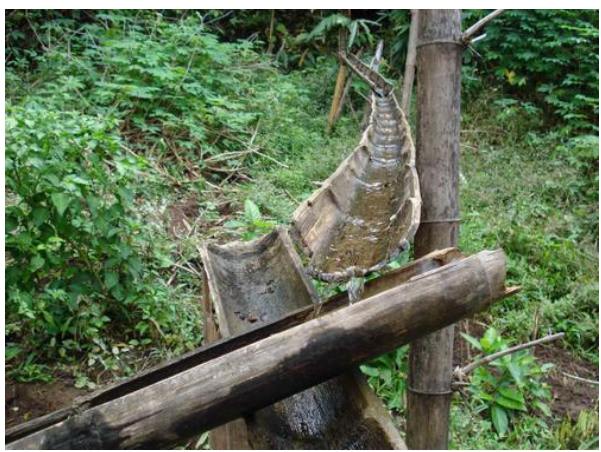

*Figure 3: The water they can collect from the spring plumped it using bamboo splits and the roots grown among the trees are their confidence to stay on. From (11)*

Time of the day is determined by a person's shadow. For example, they come for their weekly ration collection at a time when their shadow is the smallest (at noon). *They have no sense of passage of time beyond the day. They identify seasons by sounds of crickets, birds etc. Cholanaikkans foresee the transition of the seasons taking indications from the nature and its life. They listen to the chirping of crickets (beetles). They say that the kind of crickets in each season varies and so also the sound. Moreover, the presence of yellow butterflies in huge number towards the end of the monsoons is yet another placard of nature announcing the arrival of sunny days ahead.* They can also hear and smell animals at great distances. This in turn indicates their extreme endurance in auditory and olfactory body functions. It follows that they have high concentration power. They are aware of plants that can provide relief from ailments. Further investigation through interacting with them would throw more light on their traditional knowledge of medicine. *However, they do not ferment foods to produce alcohol and are unaware of intoxicating drinks (except by contamination though recent contacts).* Form this it may be inferred that they are ignorant about 'fermentation' process. Thus they may have better functioning of liver compared to a large section of modern urban or rural dwellers

## 9. Their astronomy

Absence of farming activities made sky-watching a never compelling affair for the Cholanaikkans. Hence their astronomy is minimal. They refer to Sun as '*Nyaram*' (close to *Nhayar* in Malayalam; which means Sunday) or *'Dinga'* and Moon as '*Thinkam*' (*Thinkal* in Malayalam represents Moon).They are very happy when the Sun is overhead as they feel being protected by the Sun. They regard Sun with reverence and bury their dead with the head towards West indicating the end of life as opposed to birth. They had a vague memory of eclipse. They remember that the sun can turn blue. One celestial feature they know well is shooting stars which they call *Katui* and to them it resembles ambers of the fire sent by the gods. They consider shooting as gods and they call them '*ChootuPaayuka'*. 'Choot' in Malayalam means 'Dry cocoanut Leaves, which is used for making Fire torches'; 'Payuka' means 'to run fast'. Thus they correlate with the phrase with comets or shooting stars. They know stars as *Koram* and *Udumbam*. They do not identify the stars and know no star patterns – in the planetarium when the night sky was shown to them, they smiled and walked out. They also do not know the fact that the sun rises at different points in the horizon over the year and hence do not associate it with seasons.

## 10. Linguistic studies

The most extensive documentation of *Cholannaikan* language is done in (5). The text below is reproduced from (5). The reason for reproducing it here is that it gives a reader an opportunity to judge the similarity and contrast between the *Cholanaikkan* language and the different Dravidian languages. We have more or less directly reproduced from (5) and we apologize for any typing errors that may have inadvertently crept into our reproduction. In figure 6 we have shown their tribal school.

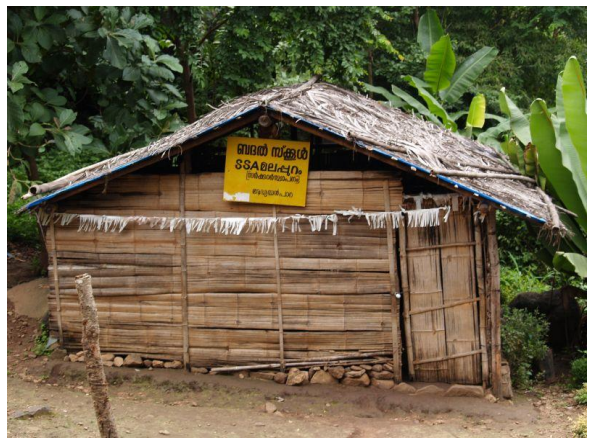

*Figure 4: Tribal School at Nilambur(wikiperida entry ion Tribals of Kerala)*

*Some linguists consider Cholanaikka language as a distinct Dravidan language in its own right, and the notion they consider the notion that it is an admixture of Tamil, Malayalam and Kannada as erroneous.*

*An interesting feature of Cholanaikkan speech is the absence of plural markers, which are used in all other Dravidian languages. Hence they use. ondumara* 'one tree' *and eed̥umara* 'two trees'. The accusative case marker is *-a*, which sometimes freely varies with *-e* and it occurs after the inflectional increment *–in.* For example both *kūsina and kūsine* mean 'child'. The instrumental case marker is – *indu*, which freely varies with *-indu*. The *–indu* marker is used more frequently. This is not related to proto Dravidian suffixes. For example, *kayttindu* means 'by knife'. The ablative case is expressed by the marker *-liddu* and it has two variants, viz., *-liddu* and *–ddu*. For example *alliddu* means 'from there' while *maraliddu* 'from the tree'.

The genitive case marker is -∅ and it freely varies with variant -e. This cannot be considered as a reflex of the proto Dravidian. *-a or *-atu. Muralidharan (4) considers this is an independent innovation in this language. For example the words *ennu* and *ennule* are used to imply 'my'. Similarly, the locative marker is –lu and it has a variant –kad̥e. –lu occurs only with non-human nouns. For example *maralu* means 'on the tree' while *ennukad̥e* 'with me'. The purposive case is marked by *-gāgi*, which freely varies with *-ga*. Hence *manegāgi* means 'for the house' can also be pronounced as *manega*.

The vocative case is marked by the marker *-ā* and it has three variants viz., *-ā*, *-e* and *-ī*. The first one occurs after nouns ending in *-annu*. And *-ē* occurs after stems ending in *-u* or *-e*, and *-ī* occurs with feminine nouns. The personal pronouns are the given in table 1;

Table 1: Personal Pronouns

| Cholanaikkan word | Meaning | Cholanaikkan word | Meaning |
|---|---|---|---|
| *naanu* | I | *nānke/nanke* | we |
| *en* | my | *Enke* | our |
| *niinu* | you | *nīnke/ninke* | you (Pl.) |
| *taan/tan* | oneself | *tanke/tānke* | themselves |
| *avnu* | that-he | *Avru* | that-they |
| *ivnnu* | this-he | *Ivru* | this-they |
| *ave* | that-she | *Ive* | this-she |
| *adu* | that-it | *idu* | this-it |

In general the personal pronouns agree with those of Malayalam. The interrogative forms of Cholanaikka are listed in table 2;

Table 2: Interrogative forms

| Cholanaikkan word | Meaning | Cholanaikkan word | Meaning |
|---|---|---|---|
| *Aanu* | who | *eennei* | how many |
| *Evnu* | which man | *eve* | which woman |
| *Edu* | which thing | *evru* | who |
| *Elli* | where | *enda* | what kind |
| *Etteku* | which side | *endu/ētteku* | when /now |
| *Ende* | in which manner | | |

The gender system agrees with other south Dravidian languages except Toda. The masculine gender markers *-nu*, and *-anu* can be connected with proto Dravidian suffixes.

The feminine gender markers are –*itti, -i, -e, -ci, -atti, -iti* and –∅. The markers *-tti* and *-e* are reconstructed to south Dravidian Suffixes. For example *-atti* is attested in all south Dravidian Languages and Telugu. The marker *-cci* is retained by Tamil, Kodagu and Kannada. In other cases, they prefix *gandu* and *ennu* to denote male and female gender respectively. Hence *eṇṇukūsu* is a female child while *gaṇḍ̥āṭu* means a male goat.

The future tense marker –mu found in this language is not found in any other South Dravindan language. This is considered as an independent innovation in this language. After future tense markers –*um* and –*mu* there is no <u>–*adu*</u> (personal termination) ending. In all Dravidian languages except Malayalam there is pronominal termination.

A word typically has Verbal base + tense marker + personal termination (va + nt + aann 'came' in Tamil). However, Muralidharan (1988) points out that in this language, the common pronominal termination *–adu* is used for all persons of past and present tenses. The form *–∅* personal termination, is used for the future tense. This is a special feature of this language.

The negative existential is denoted by the addition of auxiliary *–illa* to the infinitive form of the main verb and verbal noun forms in this language. Hence *baralilla* implies 'won't come' while *tinnadilla* implies. 'Won't eat'. The causative markers *-picc-* and the permissive marker *-aku* are also unique features of this language. Hence *noodₒpiccum* means 'will cause (someone) to see' and *nillaku* '(one) may stand' while *caadₒaku* means '(one) may jump'

The hortative form is expressed by the marker -il as in Kannada. Hence *nadₒli* means 'let (someone) plant' while *tereli means* 'left (someone) open'. Some additional peculiar lexical items found in this language are given in table 3.

Table 3: Peculiar Lexical Items

| Cholanaikkan word | Meaning | Cholanaikkan word | Meaning |
|---|---|---|---|
| *iṇdₒreci* | Wife | *iidₒubooyi* | 'penis' |
| *Eme* | Frog's croak | *eru* | join |
| Aviḷikuusu | twin | udiletaadddi | moustache |
| divvenumkalu | crescent moon | cikku | hiccough |
| cinnoonnu | mole | karse | lungs |
| Cuvale | earthworm | kuuṭruseyṭṭi | dove |
| Kaanana | dowry | kuuyennu | friend |
| goolₒimara | banyan tree | mondu | rainbow |
| niiraadₒalu | puberty | | |

All this suggests that the language, while sharing some characteristics of the Dravidian languages, they have subtle but significant differences which may suggest a parallel growth of the language.

**Conclusion**

*Cholanaikkans* are clearly an unusual tribe with several unique characteristics that under severe strain of modernisation. It is therefore important and urgent to document all aspects of their life, language and belief systems. Taking advantage of their visit to Regional Science Centre, Kozhikode on March 3, 2016, we have attempted to record of these people to the extent possible. These are reported here.

In many ways they define a unique culture with little connection with other tribal communities of India. They have very intermittent relation with the settled communities of India. This shows up in their approach to other people and in their own base of learning and science (9, 10). Their lifestyle, language and scientific belief systems are highlight their individuality and identity. We speculate that they also live in a region rich in megaliths and may well be an Austro Asian tribe that built such megaliths. This needs to be investigated through their genetic profiling.